\documentclass[aip,reprint]{revtex4-1}
\usepackage{amsmath,amssymb}
\usepackage{mathbbol}
\usepackage{bm}
\usepackage[colorlinks]{hyperref}
\usepackage[english]{babel}
\usepackage{graphicx}

\newcommand{\sign}{\mathop {\rm sign}}
\renewcommand{\Im}{\mathop {\rm Im}}

\usepackage{color}
\newcommand{\add}[1]{{#1}}

\begin{document}

%\preprint{}

\title{Optical properties of charged excitons in two-dimensional \add{semiconductors}}



\author{M.M. Glazov}
\email[]{glazov@coherent.ioffe.ru}
\affiliation{Ioffe Institute, 194021, St. Petersburg, Russia}

%\author{\ldots}
%\affiliation{\ldots}


\date{\today}

\begin{abstract}
Strong Coulomb interaction in atomically-thin transition metal dichalcogenides makes these systems particularly promising for studies of excitonic physics. Of special interest are the manifestations of the charged excitons, also known as trions, in the optical properties of two-dimensional semiconductors. In order to describe the optical response of such a system, the exciton interaction with resident electrons should be explicitly taken into account. In this paper we demonstrate that this can be done both in the trion (essentially, few-particle) and Fermi-polaron (many-body) approaches, which produce equivalent results provided that the electron density is sufficiently low and the trion binding energy is much smaller than the exciton one. Here we consider the oscillator strengths of the optical transitions related to the charged excitons, fine structure of trions and Zeeman effect, as well as photoluminescence of trions illustrating the applicability of both few-particles and many-body  models.
\end{abstract}


\maketitle %\maketitle must follow title, authors, abstract and \pacs

\section{Introduction}\label{sec:intro}

Coulomb interaction is highly important in semiconductors.  The concept of the small-radius excitons, electrons and holes tightly bound to neighbouring lattice cites, suggested by Ya.I. Frenkel\cite{PhysRev.37.17} has been extended by G. Wannier\cite{PhysRev.52.191} and N. Mott\cite{Mott384}  who demonstrated that in a number of semiconductors the hydrogen-like large radius excitons can be formed. The large radius excitons were discovered in cuprous oxide by E.F. Gross and N.A. Karryew\cite{gross:exciton:eng} in 1952 and are actively studied since then. Excitons govern optical properties of bulk semiconductors and semiconductor nanostructures.\cite{excitons:RS,ivchenko05a,Klingshirn2012}

Shortly after discovery of large radius excitons the atomic physics analogy has been extended and the excitonic molecules, also termed as biexcitons, and charged excitons, known as trions, have been predicted.\cite{PhysRevLett.1.450} The latter three-particle complexes, negative and positive trions are formed of two identical charge carriers and an unpaired one with an opposite sign: two electrons and a hole (X$^-$) and two holes and an electron (X$^+$). They are analogues of the hydrogenic ions H$^-$ (a proton and two electrons) and H$_2^+$ (two protons and an electron). The binding energies of these excitonic complexes are quite small in bulk semiconductors. The reduction of dimensionality and transition from the bulk form of materials to their two-dimensional (2D) counterparts -- quantum wells -- results in substantial increase of the trion binding energies~\cite{PhysRevLett.49.808,Stebe:1989aa,Sergeev:2001aa} which led to observation of trions in CdTe~\cite{PhysRevLett.71.1752} and GaAs~\cite{PhysRevB.53.R1709} quantum wells and initiated extensive experimental and theoretical studies of these complexes in various semiconductor nanosystems.\cite{ivchenko05a,PhysRevLett.112.147402}

Recently emerged atomically thin semiconductors based on transition metal dichalcogenide monolayers (TMDC ML) demonstrate spectacular optical properties and enhanced Coulomb effects.\cite{Mak:2010bh,RevModPhys.90.021001} The trions have been observed in these materials as well\cite{Mak:2013lh} and their fine structure and dynamics are actively studied nowadays.\cite{Plechinger2016,Courtade:2017a,PhysRevLett.123.167401,PhysRevB.101.161402} Multivalley band structure of the TMDC MLs makes it possible to observe charged biexcitons as well.\cite{Barbone:2018aa,Chen:2018aa}

There are, however, fundamental questions related \add{to trion formation} and their manifestations in optical properties of two-dimensional semiconductors. Indeed, the trions can be formed only in the presence of resident electrons, which makes it necessary to take into account the interaction of the exciton with the Fermi-sea of charge carriers rather than with a single electron. This manybody problem turns out to be extremely involved even in the limit of high carrier density.\cite{Mahan67b,Nozieres69a} Several approaches have been applied to study the interactions between excitons and free electrons in 2D structures, including direct calculation of optical susceptibility of the structure via the equations of motion or diagrammatic treatment of the electron-exciton correlations.\cite{Hawrylak91,PSSB:PSSB317,PSSB:PSSB343,suris:correlation,Klochikhin2014310,PhysRevB.94.041410} On the other hand, the problem of exciton interacting with the Fermi-sea of electrons resembles the famous polaron problem of an electron interacting with a ionic crystalline lattice~\cite{pekar1946local,landau:pekar} or an impurity atom  immersed in a Fermi gas.\cite{Koschorreck:2012aa,Schmidt:2018aa} Thus, the concept of Fermi-polarons and dressed electron-exciton excitations has been put forward and applied to study the optical response of TMDC MLs.\cite{Sidler:2016aa,PhysRevB.95.035417,PhysRevB.98.235203}

It is not, however, fully obvious that different approaches should provide the same results. One important issue is related to the trion or Fermi-polaron oscillator strength. Another problem is related with the manifestations of the trion or Fermi-polaron fine structure caused by the complex spin-valley band-structure of TMDC MLs and magnetic field.  Also, the comparative analysis of some of the basic kinetic properties of trions and Fermi-polarons, e.g., photoluminescence, is absent. Thus, it is instructive to provide side-by-side derivation of these quantities in the two approaches: trion and Fermi-polaron and demonstrate convergence of these approaches, at least for specific parameter range, i.e., very low density of electrons, linear response regime. This paper is aimed to fill this gap.

\begin{figure*}[t]
\includegraphics[width=0.89\textwidth]{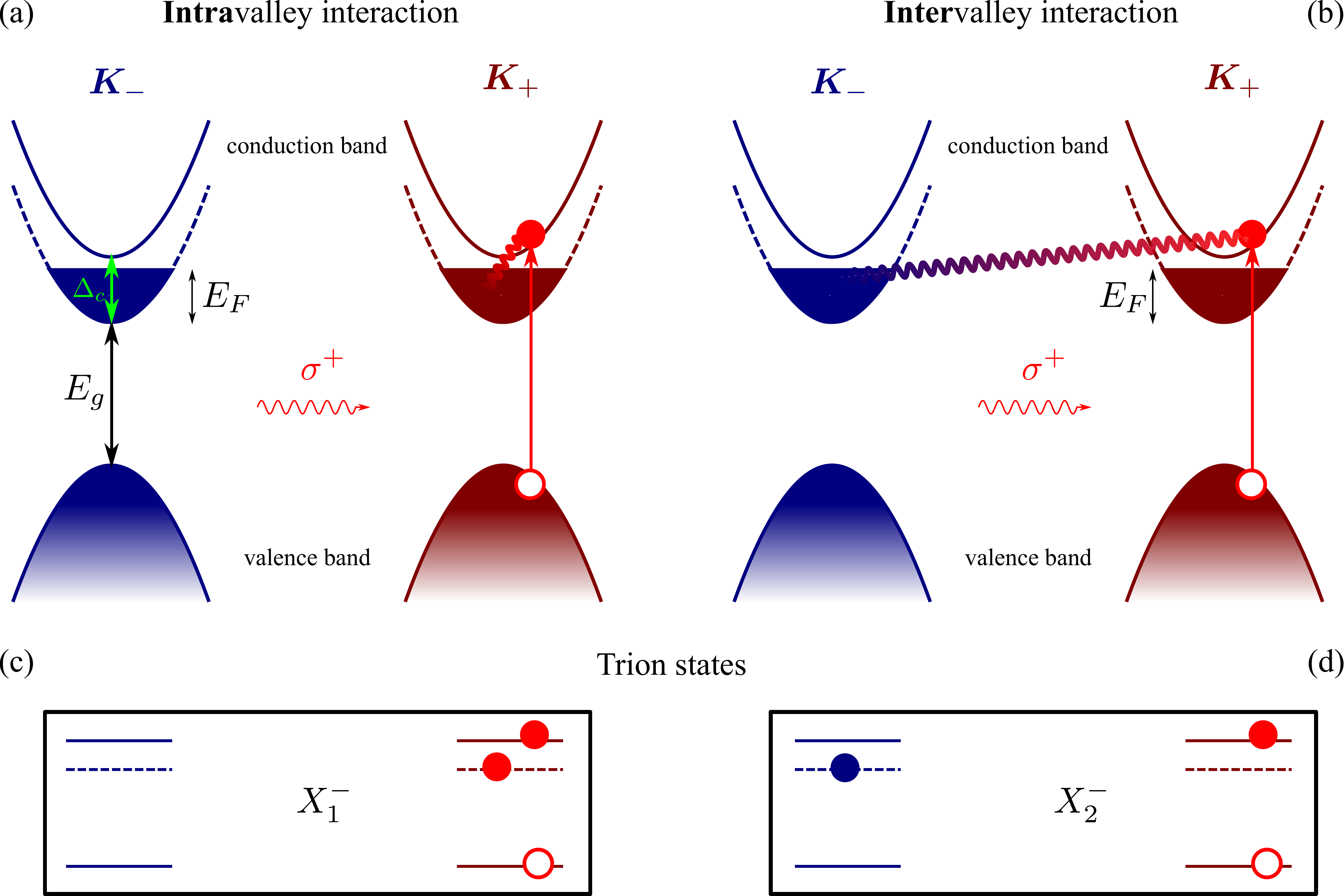}
\caption{\label{fig:bands} (a,b) Schematic band structure of lightly doped TMDC ML. Vertical arrow shows optical transition in $\sigma^+$ polarization, thick wavy line denotes the interaction of the photoexcited electron-hole complex with the Fermi-sea: (a) intravalley interaction, (b) intervalley interaction. Panels (c,d) show corresponding intra- and intervalley trions. The spin-orbit split valence subband is not shown (it is much lower in energy). The bottom conduction subbands are shown by dashed lines and assumed to be the spin-unlike with the top valence band (we consider W-based MLs). The optical transitions in $\sigma^-$ polarization involve excitation of the opposite valley $\bm K_-$.}
\end{figure*}

\section{Model}\label{sec:model}

We consider the excitonic effects in TMDC MLs within the effective mass approach, which provides simplified but physically transparent picture of the Coulomb effects in semiconductors.
The three-particle bound states of electrons and holes in TMDC ML can be described within the effective mass approach by the wavefunction\cite{Courtade:2017a,Durnev_2018}
\begin{equation}
\label{wave:trion}
\Psi_{i,j;k} = e^{\mathrm i\bm K \bm R}\varphi(\bm \rho_i,\bm \rho_j) \mathcal U_{ij}^{(2)}(\bm r_i,\bm r_j) \mathcal U_{k}^{(1)}(\bm r_k).
\end{equation}
Here the subscripts $i$ and $j$ denoted the identical charge carriers, for example two electrons $e_1$ and $e_2$ in the X$^-$ trion, and $k$ denotes the unpaired charge carrier, e.g., hole in the X$^-$ trion, $\bm r_{i,j,k}$ are the coordinates of these particles, $\bm \rho_i$ and $\bm \rho_j$ are the relative coordinates of identical particles with respect to the unpaired one, $\bm R$ is the center of mass position; hereafter the normalization area is set to unity. In Eq.~\eqref{wave:trion} $\mathcal U_{ij}^{(2)}(\bm r_i,\bm r_j)$ is the Bloch function of the electron pair (in the case of X$^-$ trion) and $ \mathcal U_{k}^{(1)}(\bm r_k)$ is the Bloch function of the unpaired hole, $\bm K$ is the wavevector of the trion translational motion, and $\varphi(\bm \rho_i,\bm \rho_j)$ is the smooth envelope of the relative motion in the trion. In what follows we consider only the ground state of the trion, focussing on the X$^-$ case. Thus, the envelope function $\varphi(\bm \rho_i,\bm \rho_j)$ is symmetric under permutations of electrons, $\bm \rho_1 \leftrightarrow\bm \rho_2$, while the two-electron Bloch function $\mathcal U_{ij}^{(2)}(\bm r_i,\bm r_j)= \mathcal U_{ij}^{(2)}(\bm r_j,\bm r_i)$ is odd and ensures the antisymmetry of the total wavefunction.\cite{Courtade:2017a}

Figure~\ref{fig:bands} illustrates the band structure of the TMDCs monolayers with two valleys $\bm K_\pm$ and the spin-orbit splitting in the conduction band; the spin-orbit splitting in the valence band is large and is not shown. We consider the W-based 2D TMDC where the spins of the bottom conduction band and top valence band are opposite,\cite{2053-1583-2-2-022001,Durnev_2018,Courtade:2017a,Wang:2017b} so the optical transition takes place to the excited spin subband of the conduction band as shown in Fig.~\ref{fig:bands}(a,b). In the presence of doping with electron Fermi energy $E_F$ being  much smaller than the conduction band spin-orbit splitting $\Delta_c$, the photocreated exciton can interact both with the electron gas in the same [$\bm K_+$ for $\sigma^+$ excitation, Fig.~\ref{fig:bands}(a)] or in the opposite [$\bm K_-$, Fig.~\ref{fig:bands}(b)] valley. In the trion picture, the exciton picks up the electron from the corresponding Fermi-sea and forms the intra- and intervalley trions shown in Fig.~\ref{fig:bands}(c) and (d), respectively.\cite{Yu:2014fk-1,Yu30122014,Courtade:2017a} Most of the results are also relevant for the Mo-based TMDC MLs, where the optical transitions at the normal incidence of radiation involve the bottom conduction subbands. In this situation, under moderate doping, only the intervalley interaction similar to that shown in Fig.~\ref{fig:bands}(b,d) is important, which makes the trion fine structure quite simple. However, a complication arises due to the fact that the photoelectron is excited to the already partially occupied band and the state filling effects related to the Pauli-blocking could be of importance. The main conclusions of this work do not largely depend on the band structure model.

In this section and in Sec.~\ref{sec:osc} we disregard, for transparency of presentation, the spin/valley structure of the Bloch functions. We address the trion fine structure in Sec.~\ref{sec:fine}.

The smooth envelope function, $\varphi(\bm \rho_1,\bm \rho_2)$ in Eq.~\eqref{wave:trion},  can be determined from the solution of the corresponding Schr\"odinger equation either variationally\cite{PhysRevB.88.045318,Courtade:2017a} or using exact analytical\cite{PhysRevLett.114.107401,PhysRevB.100.245201} or numerical\cite{2019arXiv191204873F} methods. 
In the Fermi-polaron picture, however, it is instructive to further simplify the model and consider the exciton as a rigid particle, which attracts the electron by short-range forces\cite{suris:correlation,PhysRevB.95.035417}, see Ref.~\onlinecite{2019arXiv191204873F} for detailed analysis and extensions of the model. To that end, we present the exciton-electron scattering amplitude in the form
\begin{multline}
\label{scattering:ampl}
T(\varepsilon) 
=\frac{V_0}{1+ \mathcal D V_0 \left[ \ln{\left|\frac{\tilde E - \varepsilon}{\varepsilon}\right|}+\mathrm i \pi\theta(\varepsilon)\right]}\\
= \frac{1}{\mathcal D }\frac{1}{ \ln{\left[\frac{ \varepsilon - \tilde E}{\varepsilon}\exp{\left(\frac{1}{\mathcal D V_0} \right)}\right]}}{.}
\end{multline}
Here $V_0$ is the bare matrix element of the exciton-electron scattering being short-range in the model of the rigid exciton, $\varepsilon$ is is the kinetic energy of the relative electron-exciton motion, $\mathcal D=m/(2\pi\hbar^2)$ is the reduced electron-exciton density of states with $m=m_em_x/m_{tr}$ being the reduced mass ($m_e$ is the electron effective mass, $m_h$ is the hole mass, $m_x=m_e+m_h$ is the exciton mass, and $m_{tr}=2m_e+m_h$ is the trion mass), $\theta(x)$ is the Heaviside $\theta$-function, $\theta(x)=1$ for $x>0$ and $0$ otherwise, $\tilde E$ is the cut-off energy, $\varepsilon \ll \tilde E$, which naturally arises in the 2D short-range scattering problem.  The cut-off energy introduced in Eq.~\eqref{scattering:ampl} is on the order of the exciton binding energy $E_x$: For $\varepsilon \ll E_x$ the rigid  exciton model is valid, but at $\varepsilon \gtrsim E_x$ the internal structure of the exciton should be taken into account. At $\varepsilon>0$ the scattering amplitude contains both real and imaginary parts with the latter responsible for the real scattering processes, while at $\varepsilon<0$ the amplitude $T(\varepsilon)$ is real. In derivation of Eq.~\eqref{scattering:ampl} the phase-space filling effects are disregarded, this is just a solution of a two-body ``electron+exciton'' problem.

We are interested in the situation where the electron-exciton interaction is attractive,  $V_0<0$. Thus, $T(\varepsilon)$ has a pole at a certain negative $\varepsilon$ corresponding to the bound trion state.\cite{ll3_eng,suris:correlation,PhysRevA.84.033607} We introduce the trion binding energy from the condition $T^{-1}(-E_{tr})=0$:
\begin{equation}
\label{trion:binding}
E_{tr} = \tilde E \exp{\left(\frac{1}{\mathcal D V_0}\right)} \ll \tilde E \sim E_x,
\end{equation} and  recast Eq.~\eqref{scattering:ampl}  in the alternative form
\begin{equation}
\label{scattering:ampl:1}
T(\varepsilon) = \frac{1}{\mathcal D}\frac{1}{\ln{\left(-\frac{E_{tr}}{\varepsilon} \right)}} \approx  \frac{  \mathcal D^{-1}E_{tr}}{\varepsilon + E_{tr}}.
\end{equation}
The approximate equality holds at $\varepsilon \approx - E_{tr}$, i.e., in the vicinity of the  trion pole.
Note that the model formulated above is valid provided that $E_{tr} \ll E_x$ or $|\mathcal DV_0| \ll 1$. In this approach to the Fermi-polaron problem, the trion binding energy $E_{tr}$ is the free parameter of the model which should be taken from experiments or microscopic calculations. The relative motion bound state wavefunction reads
\begin{equation}
\label{wave:tr:bound}
\Phi(\bm \rho) \propto \mathrm K_0(\rho/a_{tr}), \quad a_{tr} = \sqrt{\frac{\hbar^2}{2m E_{tr}}},
\end{equation}
where $\bm \rho$ is the electron-exciton relative motion coordinate, $\mathrm K_0$ is the modified Bessel function (Hankel function of imaginary argument), $a_{tr}$ is the effective trion radius. We also introduce the effective exciton radius by analogy with Eq.~\eqref{wave:tr:bound}, $a_x = \hbar/\sqrt{(2mE_x)} \ll a_{tr}$.

\begin{figure}[tb]
\includegraphics[width=\linewidth]{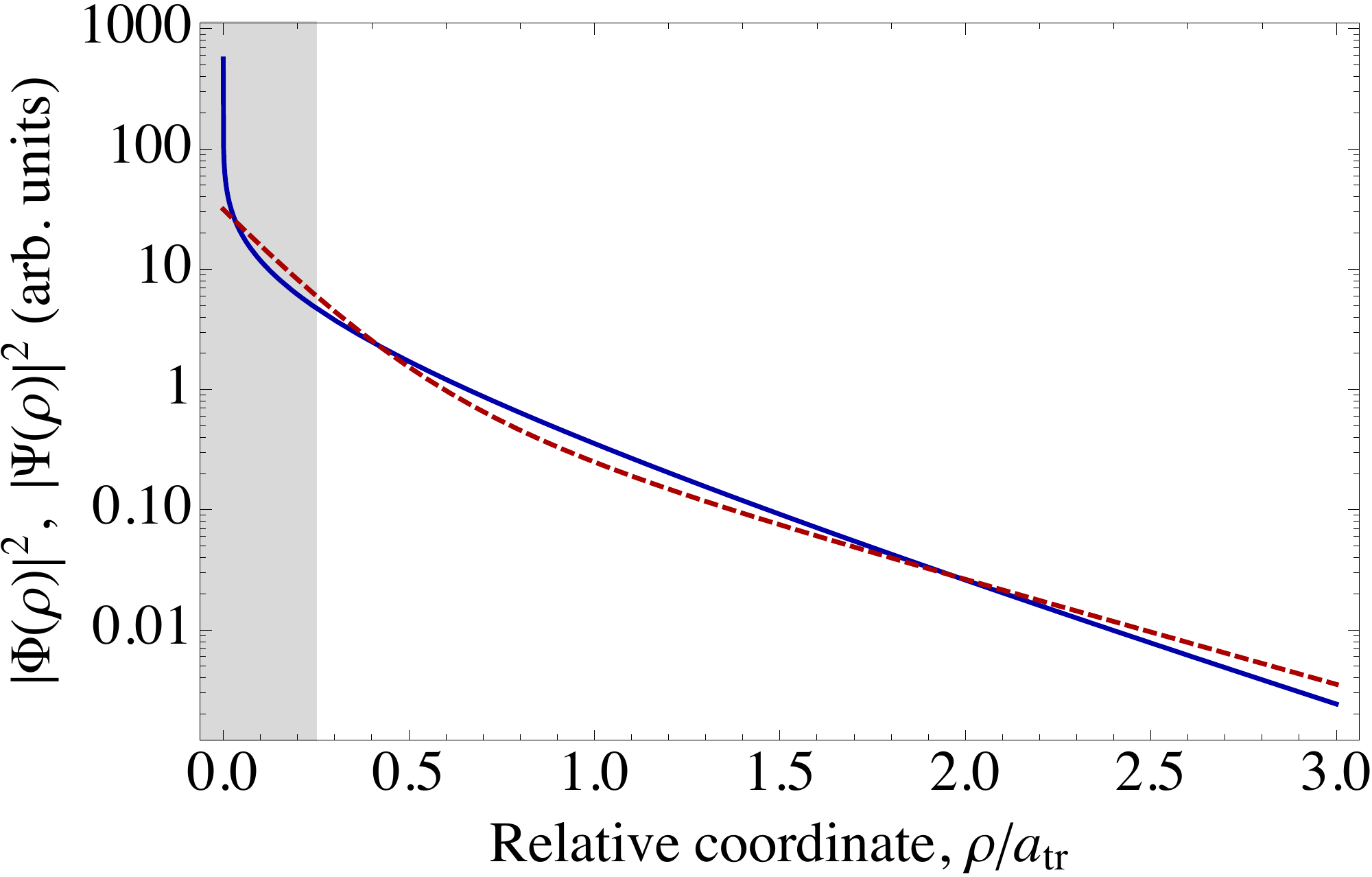}
\caption{\label{fig:functions}Probability density for exciton-electron relative motion. Blue solid line shows $|\Phi(\bm \rho)|^2$ calculated after Eq.~\eqref{wave:tr:bound} {[short-range interaction model used in our approach to  the Fermi-polaron]}. Dark red dashed line shows $ |\Psi(\bm \rho)|^2$ calculated after Eqs.~\eqref{trial} and \eqref{prop:trial} {[variational approach to  the trion wavefunction]}. Exciton radius $a_x = a_{tr}/4$. Shaded area shows the range of small $\rho \leqslant a_{x}$ where Eq.~\eqref{wave:tr:bound} is inapplicable.}
\end{figure}

It is instructive to compare the relative motion wavefunctions in the full model, Eq.~\eqref{wave:trion} and in the simplified model. For comparison, we take the envelope function in Eq.~\eqref{wave:trion} in the exponential form
\begin{equation}
\label{trial}
\varphi(\bm \rho_1,\bm \rho_2) \propto e^{-\rho_1/a_x-\rho_2/a_{tr}} + e^{-\rho_2/a_x-\rho_1/a_{tr}}, 
\end{equation}
see Refs.~\onlinecite{PhysRevB.88.045318,Courtade:2017a} for discussion of applicability of such trial function, and extract the probability density for the bound state as
\begin{equation}
\label{prop:trial}
|\Psi(\bm \rho)|^2 = \int d\bm \rho' |\varphi(\bm \rho,\bm \rho')|^2.
\end{equation}
The functions $|\Phi(\bm \rho)|^2$ and $|\Psi(\bm \rho)|^2$ are plotted in Fig.~\ref{fig:functions} and qualitatively agree with each other. Note that using more sophisticated form of the electron-hole scattering amplitude one can reproduce the results of the trion variational calculations within the scattering amplitude approach.\cite{2019arXiv191204873F}

In summary, let us highlight the relations between the system parameters where the developed approach is valid. We consider here the 2D semiconductor with free electrons. Importantly, the following hierarchy of energies should take place
\begin{subequations}
\label{conditions}
\begin{equation}
\label{cond:energ}
E_x \gg E_{tr} \gg E_F,
\end{equation}
i.e. the exciton binding energy (typically hundreds of meV) should exceed by far the trion binding energy (typically tens of meV), which, in its turn, should be much larger than the electron Fermi energy $E_F$. This inequality can be translated to the equivalent relation between the length scales
\begin{equation}
\label{cond:lengths}
a_x \ll a_{tr} \ll 1/k_F,
\end{equation}
where $k_F = \sqrt{2m_e E_F/\hbar^2}$ with $m_e$ being the electron effective mass is the Fermi wavevector of the electron. Equation~\eqref{cond:lengths} has a transparent physical meaning: The exciton can be considered as a small rigid particle and the trion is formed by attaching the electron to this particle. Furthermore, the Coulomb binding of excitons and trions should occur on the small length scales as compared with the characteristic wavelengths of the resident electrons, in order to be able to treat the exciton interaction with Fermi-sea perturbatively. It is noteworthy, however, that due to numerical factors Eq.~\eqref{cond:lengths} provides more stringent conditions than relation between the energies, Eq.~\eqref{cond:energ}. Further, for simplicity we assume that the temperature, $T$,  expressed in the units of energy is low 
\begin{equation}
\label{cond:temp}
k_B T \ll E_F,
\end{equation}
\end{subequations}
i.e., the electrons are degenerate and thermal excitations can be neglected. This condition is, in general, not mandatory, and can be easily relaxed. Only quantitative changes are expected for $k_B T \ll E_{tr}$. We note that the condition $E_x \gg E_F$, Eq.~\eqref{cond:energ}, also implies  \add{that} the electron-electron interactions in the Fermi-sea are parametrically strong. \add{We disregard the effects of Wigner crystallization of electrons because typically there is a parameter range where such collective effects are unimportant even at $E_F \ll E_x$.\cite{WC} Therefore, we assume} that the electrons can be still treated as weakly-interacting quasi-particles with all Coulomb effects included in renormalized values of their parameters (Fermi energy, effective mass). For instance, Refs.~\onlinecite{Hawrylak91,PhysRevB.94.041410} consider the crossover between different regimes of the exciton-electron interactions with variation of electron density. We also note that due to the conditions $E_{tr} \gg E_F$ one can, at least in the first approximation, disregard the state-filling effects in the case of Mo-based 2D TMDCs where the optical transition involves the trion formation in the already occupied valley.

\section{Oscillator strength}\label{sec:osc}

The key parameter controlling the optical response of the excitonic complexes in semiconductors is the oscillator strength, which describes the efficiency of the light-matter interaction. In this section we calculate the oscillator strength both in the trion and Fermi-polaron approaches and compare the results.

The resonant trion excitation can be considered as a process where (i) an exciton is created in the virtual intermediate state and (ii) the exciton picks up an electron from the Fermi-sea to form a trion. Thus, a finite density of resident electrons is needed to make this process possible.

\subsection{Fermi-polaron approach}\label{sec:osc:FP}

In the Fermi-polaron approach the optical response function can be readily expressed via the  exciton Greens function\footnote{The full Greens function of excitons depends, generally, on two wavevectors $\mathcal G_x(\varepsilon; \bm k,\bm k')$, since our system is translationally invariant one can put $\bm k'=\bm k$. The full Greens function will be needed in Sec.~\ref{sec:PL:FP}, where we take into account exciton-electron and exciton-phonon interactions simultaneously.} 
\begin{equation}
\label{exciton:Greens}
\mathcal G_x(\varepsilon;\bm k) = \frac{1}{\varepsilon - E_{\bm k} - \Sigma(\varepsilon;\bm k) + \mathrm i \Gamma}.
\end{equation}
Here $E_{\bm k}$ is the exciton dispersion in TMDC ML plane, $\bm k$ is the exciton in-plane wavevector, $\Gamma$ is the exciton damping rate caused, e.g., by the exciton-phonon interaction, inhomogeneous broadening, etc.,\footnote{Generally, $\Gamma$ also includes the radiative damping of the exciton, which, however, needs to be found self-consistently from the solution of Maxwell equations with susceptibility~Eq.~\eqref{Pi}, see Refs.~\onlinecite{ivchenko05a,PhysRevLett.123.067401}} $\Sigma(\varepsilon;\bm k)$ is the exciton-self energy resulting from the interaction with resident electrons. We take it in the simplest form using Eq.~\eqref{scattering:ampl} (see Appendix~\ref{sec:app:self} for more detailed discussion):
\begin{equation}
\label{exciton:self}
\Sigma(\varepsilon;\bm k) = T(\varepsilon) N_e,
\end{equation}
where $N_e$ is the electron density.
Corresponding optical susceptibility in a given circular polarization at the normal incidence of radiation can be written as 
\begin{equation}
\label{Pi}
\Pi(\omega) =  f_x \mathcal G_x(\hbar\omega - E_g + E_x;0), \quad f_x = |\mathfrak M_r|^2 |\varphi_x(0)|^2  
\end{equation}
Here $f_x$ is the effective exciton oscillator strength, $\mathfrak M_r$ is the matrix element of the interband transition (per photon), $\varphi_x(\rho)$ is the exciton relative motion envelope function, and $E_g$ is the band gap, see Refs.~\onlinecite{glazov2014exciton,PhysRevLett.123.067401} for details. In this section we disregard spin and valley fine structure of the trion, the role of these intrinsic degrees of freedom is  discussed in detail in Sec.~\ref{sec:fine}.

In the vicinity of the exciton resonance where $\hbar\omega \approx E_g - E_x$, the self-energy is small and excitonic states are almost unaffected by the electron gas in respect of the exciton binding energy and wavefunction, see, however, more details below and Eq.~\eqref{coupled:Gr:1} for the analysis of the oscillator strength:\cite{Glazov:2018aa}
\begin{equation}
\label{Pi:exc:approx}
\Pi(\omega) \approx \frac{f_x}{\hbar\omega - E_g + E_x + \mathrm i \Gamma + \Sigma(\hbar\omega - E_g + E_x)}.
\end{equation}
The main important effect here is the exciton damping induced by the electron-exciton scattering: Qualitatively, it follows from Eq.~\eqref{scattering:ampl} where $T(\varepsilon)$ has an imaginary part at $\varepsilon>0$ responsible for the scattering. Quantitative discussion of this and related issues is beyond the scope of the paper.\cite{PhysRevB.95.035417,PhysRevB.99.125421,2020arXiv200310932C} Also, the exciton oscillator strength decreases as it is transferred to the attractive Fermi-polaron (trion), see below. 
The resonance in $\Pi(\omega)$ at $\hbar\omega \approx E_g - E_x$, Eq.~\eqref{Pi:exc:approx} is termed as the repulsive Fermi-polaron, see below.

Importantly, due to $\Sigma \ne 0$, particularly, due to the pole in $\Sigma(\varepsilon)$ at $\varepsilon = -E_{tr}$ another resonance -- termed as the attractive Fermi-polaron -- appears in the susceptibility at $\hbar\omega \approx E_g - E_x - E_{tr}$. Indeed, making use of approximate Eq.~\eqref{scattering:ampl:1} we arrive at [cf. Refs.~\onlinecite{suris:correlation,Sidler:2016aa,PhysRevB.95.035417}]
\begin{equation}
\label{Pi:trion}
\Pi(\omega) \approx \frac{f_{tr}}{\hbar\omega - E_g + E_x + E_{tr} + N_e/\mathcal D + \mathrm i \Gamma N_e /\mathcal D },
\end{equation}
where the effective oscillator strength 
\begin{equation}
\label{ft}
f_{tr} = \frac{N_e}{\mathcal D E_{tr}} |\mathfrak M_r|^2 |\varphi_x(0)|^2 =4\pi N_e a_{tr}^2 f_x,
\end{equation} 
and $f_x$ is introduced in Eq.~\eqref{Pi} and corresponds to the absence of doping.
Thus, part of the exciton oscillator strength is shuffled towards the Fermi-polaron peak. The peak position is at $\hbar\omega = E_g - E_x - E_{tr} -N_e/\mathcal D$. The  shift of the peak with respect to the trion energy ($E_g - E_x - E_{tr}$) is proportional to the electron Fermi energy. Namely, the quantity $\delta = N_e/\mathcal D$  can be recast as
\begin{equation}
\label{shift}
\delta = E_F \frac{m_e}{m} = E_F \frac{m_{tr}}{m_x}.
\end{equation}
We recall that here $m_{tr} = 2m_e+m_h$ is the trion translational mass, and $m_x=m_e+m_h$ is the exciton translational mass. This shift is assumed to be small, $\delta \ll E_{tr}$, cf. Eq.~\eqref{cond:energ}, otherwise the  form of the self-energy used here is insufficient, see Appendix~\ref{sec:app:self} for details.

It is instructive to introduce, based on the considerations above, even more simplified model of the Fermi-polaron. To that end we use approximate form of the scattering amplitude~\eqref{scattering:ampl:1} across the whole relevant energy range and present the exciton Greens function in the form
\begin{equation}
\label{coupled:Gr}
\mathcal G_x(\varepsilon) =\frac{1}{\varepsilon + \mathrm i \Gamma - \dfrac{N_e \mathcal D^{-1} E_{tr}}{\varepsilon + E_{tr}+ \mathrm i \gamma}}.
\end{equation}
To shorten the notations and for simplicity we put $\bm k=0$, but for generality we introduced the trion damping $\gamma$. The Greens function~\eqref{coupled:Gr} describes two coupled oscillators: One describes the exciton and another one describes the trion. The parameter $g=\sqrt{N_e \mathcal D^{-1} E_{tr}}\sim \sqrt{E_F E_{tr}}$ plays a role of the coupling constant. We consider the regime where $g \ll E_{tr}$ (analog of the weak coupling, otherwise the simplifications behind the model make it inapplicable) and recast Eq.~\eqref{coupled:Gr} in the form
\begin{equation}
\label{coupled:Gr:1}
\mathcal G_x(\varepsilon) \approx\frac{1-\mathcal D^{-1} N_e/E_{tr}}{\varepsilon - {N_e \mathcal D^{-1}}+ \mathrm i \Gamma } + \frac{\mathcal D^{-1} N_e/E_{tr}}{{\varepsilon + E_{tr} + N_e \mathcal D^{-1}+ \mathrm i \gamma}}.
\end{equation}
Equation~\eqref{coupled:Gr:1} makes it possible to introduce the notions of the \emph{attractive} and \emph{repulsive} Fermi polarons as the poles of $\mathcal G_x$ at $\varepsilon\approx - E_{tr}$ (in the vicinity of the trion resonance) $\varepsilon \approx 0$ (in the vicinity of the exciton resonance). The attractive polaron state stems from the bound trion and corresponds to the exciton strongly correlated with the resident electrons. The repulsive polaron states describe the continuum-like exciton-electrons states, i.e., exciton state perturbed by the Fermi-sea of electrons.

\subsection{Trion approach}\label{sec:osc:tr}

Now let us calculate the oscillator strength in the trion approach. For rigorous calculation one has to take into account the presence of the Fermi-sea explicitly. It can be conveniently done in the secondary quantization approach.

Light-matter coupling Hamiltonian describing optical transitions at the normal incidence of radiation reads
\begin{equation}
\label{rad:Ham}
\mathcal H_{\rm rad} = \mathfrak M_r\sum_{\bm k_e,\bm k_h} a_{\bm k_e}^\dag b_{\bm k_h}^\dag \delta_{\bm k_e + \bm k_h,0} +{ \rm h.c.},
\end{equation}
where the operators $a_{\bm k}$ ($a^\dag_{\bm k}$) correspond to an electron, $b_{\bm k}$ ($b^\dag_{\bm k}$) correspond to the hole, $\bm k_e$ and $\bm k_h$ are the in-plane wavevectors of the electron and hole, and, as above, we disregard the spin and valley structure of the electronic states (it is considered in detail in Sec.~\ref{sec:fine}).

It is convenient to calculate the matrix element of the exciton optical generation. Within the secondary quantization approach the exciton wavefunction can be written as\cite{ivchenko05a}
\begin{equation}
\label{initial:X}
|X\rangle = \sum_{\bm k_e,\bm k_h} F_x(\bm k_e, \bm k_h) a^\dag_{\bm k_e} b^\dag_{\bm k_h}|vac\rangle,
\end{equation}
where $|vac\rangle$ is the state of the 2D crystal with empty conduction and filled valence bands and $F_x(\bm k_e, \bm k_h)$ is the exciton envelope function in the $\bm k$-space. The matrix element of the optical transition to the exciton state reads
\begin{equation}
\label{rec:X}
\langle X|\mathcal H_{\rm rad}|vac\rangle =  \mathfrak M_r \sum_{\bm k} F^*(\bm k, -\bm k) = \mathfrak M_r \varphi_x^*(0).
\end{equation}
In the last equation we took into account the relation $\varphi_x(\bm \rho) = \sum_{\bm k} F(\bm k, - \bm k) \exp{(-\mathrm i \bm k \bm \rho)}$. Thus, effective oscillator strength of the exciton is given by
\begin{equation}
\label{osc:X}
f_x = \left|\mathfrak M_r\right|^2 |\varphi_x(0)|^2,
\end{equation}
in full agreement with Eq.~\eqref{Pi}.

Let us now consider the $X^-$ trion. Its wavefunction in the $\bm k$ space can be written via the Fourier transform of Eq.~\eqref{wave:trion}
\begin{equation}
\label{initial:ra}
|T\rangle = \sum_{\bm k_1,\bm k_2, \bm k_h} F_{tr}(\bm k_1, \bm k_2, \bm k_h)  a_{\bm k_1}^\dag  \tilde a_{\bm k_2}^\dag b_{\bm k_h}^\dag |vac\rangle.
\end{equation}
We used $\tilde a_{\bm k_2}^\dag$ to denote the creation operator of one of the electrons to highlight that it is in the different spin or valley state as compared with another electron.  For the trion to be formed an electron with the wavevector $\bm k_e$ should be present in the system, thus, the initial state is
\begin{equation}
\label{final:radiative}
|e\rangle =  \tilde a_{\bm k_e}^\dag|vac\rangle.
\end{equation}
\add{At small electron densities where $E_F \ll E_{tr}$ and, correspondingly, $k_F \ll a_{tr}^{-1}$ the effects of Pauli blocking in the final (trion) state can be disregarded.}

Neglecting, as before, the photon momentum we calculate the matrix element of the Hamiltonian~\eqref{rad:Ham}  and arrive at in agreement with Refs.~\onlinecite{PhysRevB.58.9926,PhysRevB.62.8232},
\begin{multline}
\label{me:rad}
\langle T|\mathcal H_{\rm rad} |e\rangle = \delta_{\bm k_e, \bm K}\mathfrak M_r \sum_{\bm k_2, \bm k_h} F^*(-\bm k_h,\bm k_2,\bm k_h) \\
= \delta_{\bm k_e, \bm K} \mathfrak M_r \int  \varphi^*(0, \bm \rho) e^{\mathrm i \bm K \bm \rho} d\bm \rho. 
\end{multline}
We stress that due to the momentum conservation law the in-plane wavevector of the electron equals to the wavevector of the trion translational motion, $\bm k_e = \bm K$.

The TMDC ML susceptibility in the vicinity of the trion resonance can be evaluated taking into account all possible initial states for electrons and, correspondingly, all possible wavevectors of the trions in the final state. Making use of the Fermi's golden rule we recast the imaginary part of the susceptibility in the form
\begin{multline}
\label{Pi:trion}
-\Im\{\Pi(\omega)\} = \sum_{\bm K} |\mathfrak M_r|^2 n_{\bm K}  
\left|\int d\bm \rho \varphi(0,\bm \rho) \exp{(-\mathrm i \bm K \bm \rho)} \right|^2 \\ \times
\frac{\gamma}{(\hbar\omega -E_g + E_x +E_{tr} - \delta_K)^2 + \gamma^2}
\end{multline}
with $n_{\bm K}$ being the electron distribution function, $N_e = \sum_{\bm K} n_{\bm K}$ and
\begin{equation}
\label{delta:k}
\delta_K = \frac{\hbar^2 K^2}{2 m_e} \frac{m_x}{m_{tr}}.
\end{equation}
This quantity takes into account the energy and momentum conservation in the process of picking the electron from the Fermi sea and forming the trion. Equation~\eqref{Pi:trion} is in agreement with Ref.~\onlinecite{PSSB:PSSB317}.

Neglecting the trion dispersion and assuming, similarly to Sec.~\ref{sec:osc:FP} [Eq.~\eqref{cond:energ}], that the electron Fermi energy is much smaller than the trion binding energy
we arrive at\footnote{Strictly speaking this result is valid at $E_F\ll \gamma$, otherwise the omission of $\delta_K$ in the resonant denominator is not justified.} 
\begin{multline}
\label{ftt:1}
f_{t} = \left|\mathfrak M_r\right|^2 \sum_{\bm K} n_{\bm K} \left|\int d\bm \rho \varphi(0,\bm \rho) \exp{(-\mathrm i \bm K \bm \rho)} \right|^2 \\
\approx N_{e} \left|\mathfrak M_r\right|^2 \left|\int \varphi(0,\bm \rho) d\bm \rho \right|^2,
\end{multline}
where in the latter approximate equation we have made a replacement $\exp{(-\mathrm i \bm K \bm \rho)}\to 1$ valid at very low electron densities [$K\sim k_F \ll a_{tr}^{-1}$, Eq.~\eqref{cond:lengths}]. Strictly speaking, the possibility to neglect $\delta_K$ in the denominator is possible if, in addition to Eqs.~\eqref{conditions}, we assume that $E_F \lesssim \gamma$, i.e., if the broadening of the trion line is sufficiently large.

\begin{figure}[t]
\includegraphics[width=\linewidth]{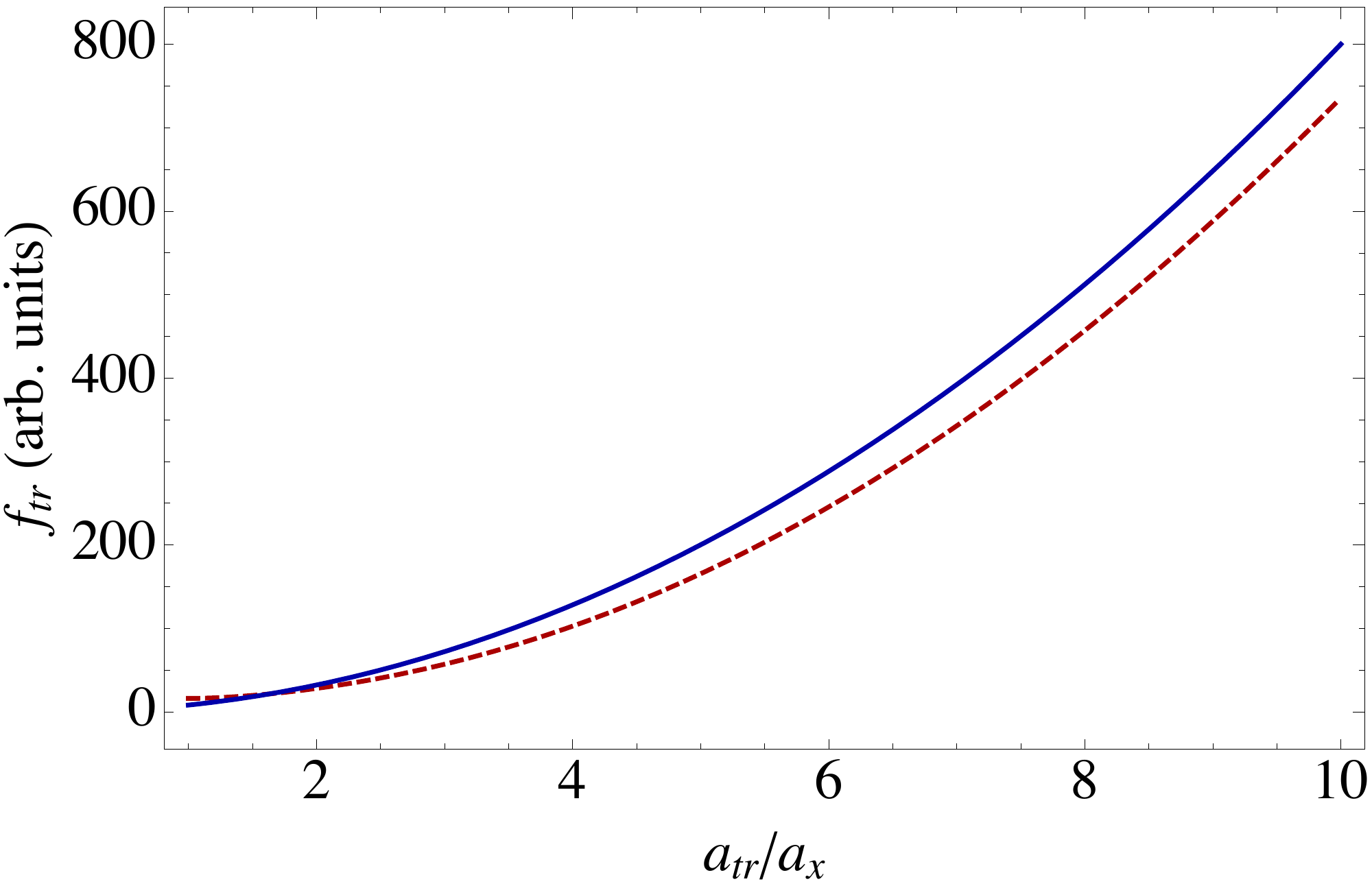}
\caption{\label{fig:osc} Effective trion oscillator strength as a function of the ratio of trion and exciton radii. Dark red dashed line shows the results of the calculation after Eq.~\eqref{ftt:2}, blue solid line shows $a_{tr}^2/a_x^2$ asymptotics.}
\end{figure}

To provide a link with the Fermi-polaron approach we use the trial function~\eqref{trial} and evaluate the trion oscillator strength from Eq.~\eqref{ftt:1} with the result:
\begin{equation}
\label{ftt:2}
f_{tr} = N_e \left|\mathfrak M_r\right|^2\frac{(a_x^2+a_{tr}^2)^2}{{\frac{a_x^2a_{tr}^2}{8}+\frac{2a_x^4a_{tr}^4}{(a_x+a_{tr})^4}}} \approx {4\pi} N_e a_{tr}^2 f_x.
%,\quad a_{tr} \gg a_x.
\end{equation}
The last approximate equality holds at $a_{tr} \gg a_x$, Eq.~\eqref{cond:lengths}, and to derive it we used the hydrogenic form of the exciton envelope function $\varphi_x(\rho) = \sqrt{2/\pi a_x^2} \exp{(-\rho/a_x)}$ with the same exciton radius.
Noteworthy, Eq.~\eqref{ft} derived in the Fermi-polaron approach and Eq.~\eqref{ftt:2} derived in the trion approach  agree  at $a_{tr} \gg a_x$. At a fixed electron density the trion oscillator strength scales as $(a_{tr}/a_x)^2$, see Fig.~\ref{fig:osc}.

We stress that the agreement of Eqs.~\eqref{ft} and \eqref{ftt:2} is not a coincidence. In both cases the process of virtual exciton creation by photon and subsequent binding with electron is described which corresponds to the resonant excitation in the vicinity of  the trion resonance. Qualitatively this explains the ratio $f_{tr}/f_x \sim N_e a_{tr}^2$, since the trion formation is only possible if there is an electron in the area $\sim a_{tr}^2$ in the vicinity of the exciton. This proportionality relation is  used in conventional semiconductor quantum wells to determine the electron density optically.\cite{astakhov00,astakhov02} Thus, oscillator strengths of the trion (attractive polaron) optical transitions can be calculated in any approach with the same result at small resident electron densities. 

It follows from Eqs.~\eqref{Pi:trion} and \eqref{delta:k} that the presence of electrons  broadens and shifts the trion resonance in the susceptibility as compared to its initial position at $\hbar\omega - E_g + E_x + E_{tr}$. The origin of the shift is somewhat similar to the ``polaron'' shift in Eq.~\eqref{shift}, the magnitude of the effect is, however, different. The difference is related with simplifications used here. \add{An accurate comparison requires calculation of the difference between the trion and exciton (attractive and repulsive polarons) positions in the spectra. This requires going beyond the approximate form for the scattering amplitude, Eq.~\eqref{scattering:ampl:1}, used above in the Fermi-polaron approach and  a self-consistent determination of the exciton self-energy, see Appendix~\ref{sec:app:self}.  In the trion approach, the electron-exciton and electron-trion interaction-induced renormalizations of the exciton and trion energies were neglected and should be taken into account. This is beyond the scope of this work.}

\section{Fine structure and Zeeman effect}\label{sec:fine}

In this section we address the trion and Fermi-polaron energy spectrum fine structure and the Zeeman effect in TMDC MLs. 

We start with the situation where the external magnetic field is absent. We recall that the short-range contributions to the electron-electron and electron-hole interaction split the intra- and intervalley trion states.\cite{Danovich:2017aa,Courtade:2017a} We denote the intravalley state as $X^-_1$ and the intervalley state $X^-_2$ and their binding energies (including the short-range contributions) as $E_{tr,1}$ and $E_{tr,2}$, respectively. Accordingly, the trion radii are different as well and denoted, respectively, as $a_{tr,1}$ and $a_{tr,2}$. Thus, the trions/Fermi-polarons  in tungsten-based MLs should appear as a doublet split by $|E_{tr,1}-E_{tr,2}|$ with slightly different oscillator strengths of individual peaks.

\begin{figure}[t]
\includegraphics[width=1\linewidth]{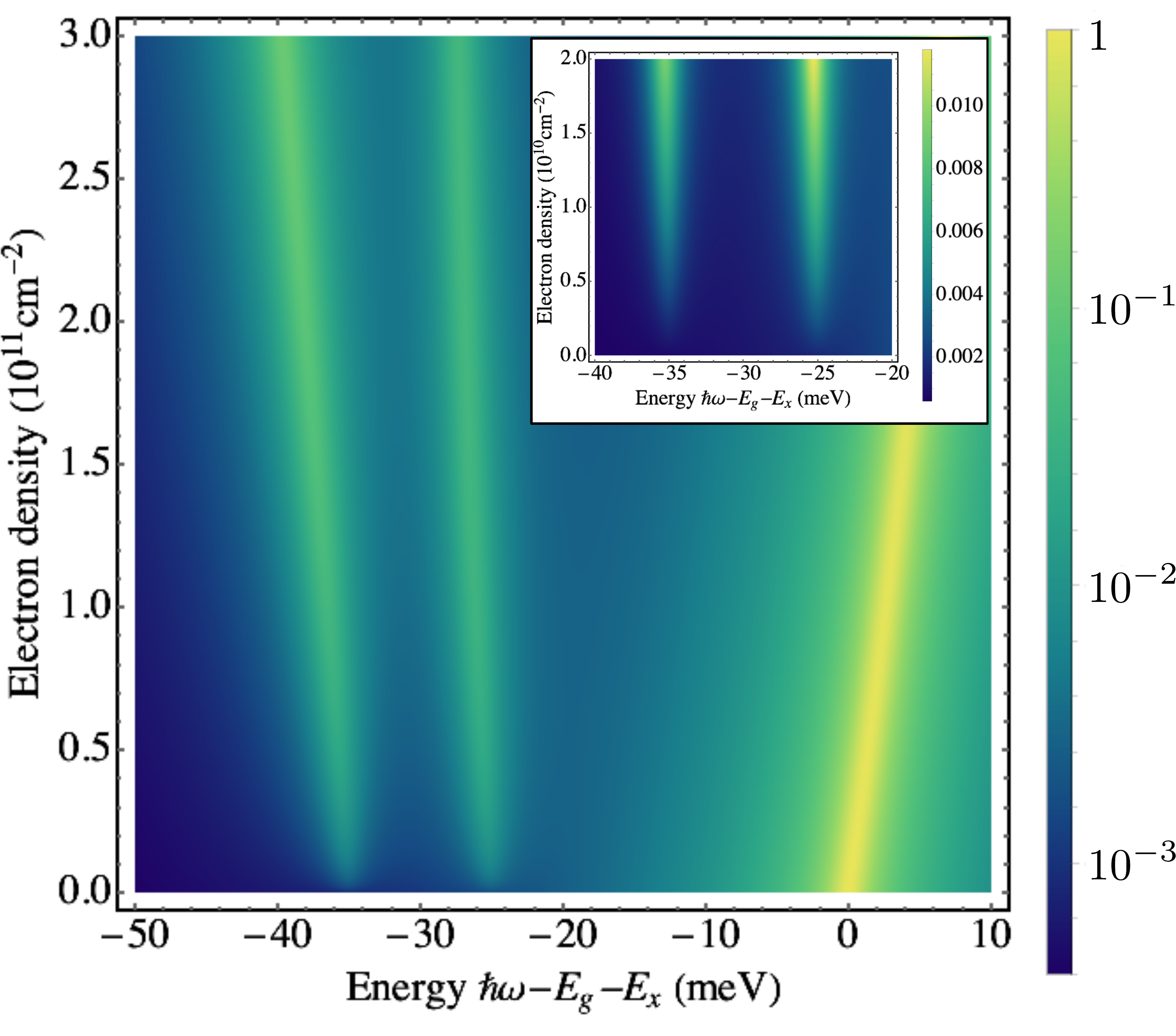}
\caption{\label{fig:Ndep} False color plot (log-scale of intensity) of the optical absorption spectrum given by $-\Im\{\Pi(\omega)\}$ calculated after Eqs.~\eqref{Pi} and \eqref{Gx:2res} in the absence of magnetic field for varied electron density (per valley). Electron and hole masses are $m_e = m_h=m_0/2$ with $m_0$ being free electron mass, $E_{tr,1}=25$~meV, $E_{tr,2}=35$~meV (exaggerated for illustrative purposes), $\gamma=\Gamma=1$~meV. Inset shows the plot in the vicinity of the trion (Fermi-polaron) resonances at low doping in the linear scale. Energy is reckoned from the exciton resonance energy at negligible doping.}
\end{figure}

Figure~\ref{fig:Ndep} demonstrates the optical absorption spectrum as a function of energy and electron density. It is calculated extending Eqs.~\eqref{Pi} and \eqref{coupled:Gr} to account for two trion states:
\begin{equation}
\label{Gx:2res}
\mathcal G_x^+(\varepsilon) =\frac{1}{\varepsilon + \mathrm i \Gamma - \dfrac{N_{e} \mathcal D^{-1} E_{tr,1}}{\varepsilon + E_{tr,1}+ \mathrm i \gamma} - \dfrac{N_{e} \mathcal D^{-1} E_{tr,2}}{\varepsilon  + E_{tr,2}+ \mathrm i \gamma}},
\end{equation}
with $N_e$ being the electron density per valley. The spectrum in Fig.~\ref{fig:Ndep} shows the strong excitonic feature (repulsive polaron) and two low-energy  trion features (attractive polarons). The appearance of the trion oscillator strength with increasing the electron density is clearly seen and it is described by the general model outlined in Sec.~\ref{sec:model}.  Note that with increasing the Fermi energy, the indirect coupling between the $X_1^-$ and $X_2^-$ appears via their interaction with excitons making redistribution of the oscillator strengths non-trivial. Here we abstain from detailed discussion of the oscillator strengths of the trion (attractive polaron) features, see Ref.~\onlinecite{zipfel:trions:marina} for more detailed analysis at low densities. Also, the polaron-like repulsion of the peaks in the optical spectrum controlled by the parameter $\delta$ in Eq.~\eqref{shift} is clearly seen. We note that the presence of electrons can also affect the band gap, exciton and trion binding energies and provide further modifications both of the absolute positions of the lines in the spectrum and also of the relative distance between the neutral and charged exciton (repulsive and attractive) polaron lines.

Let us now discuss the Zeeman effect in the presence of an external magnetic field $\bm B$ applied along the ML normal. We assume that the field is sufficiently small to disregard the orbital effects of the field both on the excitons and trions as well as on the electrons. \add{It is justified at $|e B/m_e c| \tau_e \ll 1$, where $\tau_e$ is the electron scattering time or, at finite temperature $T$ at $|e B/m_e c| \ll k_B T/\hbar$.} Thus, the magnetic field produces the Zeeman splitting of the electron and hole states lifting the Kramers degeneracy between the states in the opposite valleys, and, due to the splitting, the valley polarization of the resident electrons. For the valence band states the Zeeman effect (in the electron representation) is described by the Land\'e factor $g_v$, and the splitting equals to
\begin{subequations}
\label{Zeeman}
\begin{equation}
\label{vb}
\Delta_{Z,v} = g_v \mu_B B.
\end{equation}
It is responsible for the energy shift of the valence band in the $\bm K_+$ valley with respect to the valence band in the $\bm K_-$ valley. Note that $\Delta_{Z,v}>0$ corresponds to the $\bm K_+$ valence band top being above that of the $\bm K_-$ valence band. For the conduction band there are two spin subbands. Thus, we introduce two Land\'e factors, $g_c$ and $g_c'$, responsible for the splitting of the Kramers-degenerate pairs of the top and bottom subbands in $\bm K_\pm$ valleys, respectively:
\begin{align}
\Delta_{Z,c} = g_c \mu_B B,\label{cb:top}\\
\Delta_{Z,c}' = g_c' \mu_B B.\label{cb:bot}
\end{align}
\end{subequations}
The sign convention is the same, $\Delta_{Z,c}>0$ ($\Delta_{Z,c}'>0$) corresponds to the $\bm K_+$ state higher in energy as compared with the $\bm K_-$ state in the corresponding subband.

Since in our model the topmost subbands have the same spin as the valence band top, the splitting of the optical transitions is given by the combination of $\Delta_{Z,c}$ and $\Delta_{Z,v}$, giving rise to the bright exciton Zeeman splitting\cite{2053-1583-2-3-034002}
\begin{equation}
\label{exc:g:br}
\Delta_{Z,x} = \Delta_{Z,c} - \Delta_{Z,v} = g_x\mu_B B, 
\end{equation}
with
\[
g_x = g_c - g_v.
\]
As we also assume that $E_F \ll \Delta_c$, only the bottom conduction subbands are occupied with the electrons. Correspondingly, the Zeeman effect in the bottom subbands gives rise to the electron valley polarization ($N_{e,\bm K_\pm}$ is the electron density in the corresponding valley):
\begin{equation}
\label{Pv}
P_v = \frac{N_{e,\bm K_+} - N_{e,\bm K_-}}{N_{e,\bm K_+} + N_{e,\bm K_-}} = -\frac{1}{2} \frac{\Delta_{Z,c}'}{E_F}.
\end{equation}
In derivation of Eq.~\eqref{Pv} we assumed that $k_B T \ll E_F$, Eq.~\eqref{cond:temp}, and that $|\Delta_{Z,c}'|\leqslant 2 E_F$ ($E_F$ corresponds to the magnetic-field-less case). If the latter inequality is not satisfied, the $P_v = -\sign{\Delta_{Z,c}'}$. Here we also neglect the exchange renormalization of the electron $g$-factor.\cite{PhysRev.178.1416}

\begin{figure}[t]
\includegraphics[width=\linewidth]{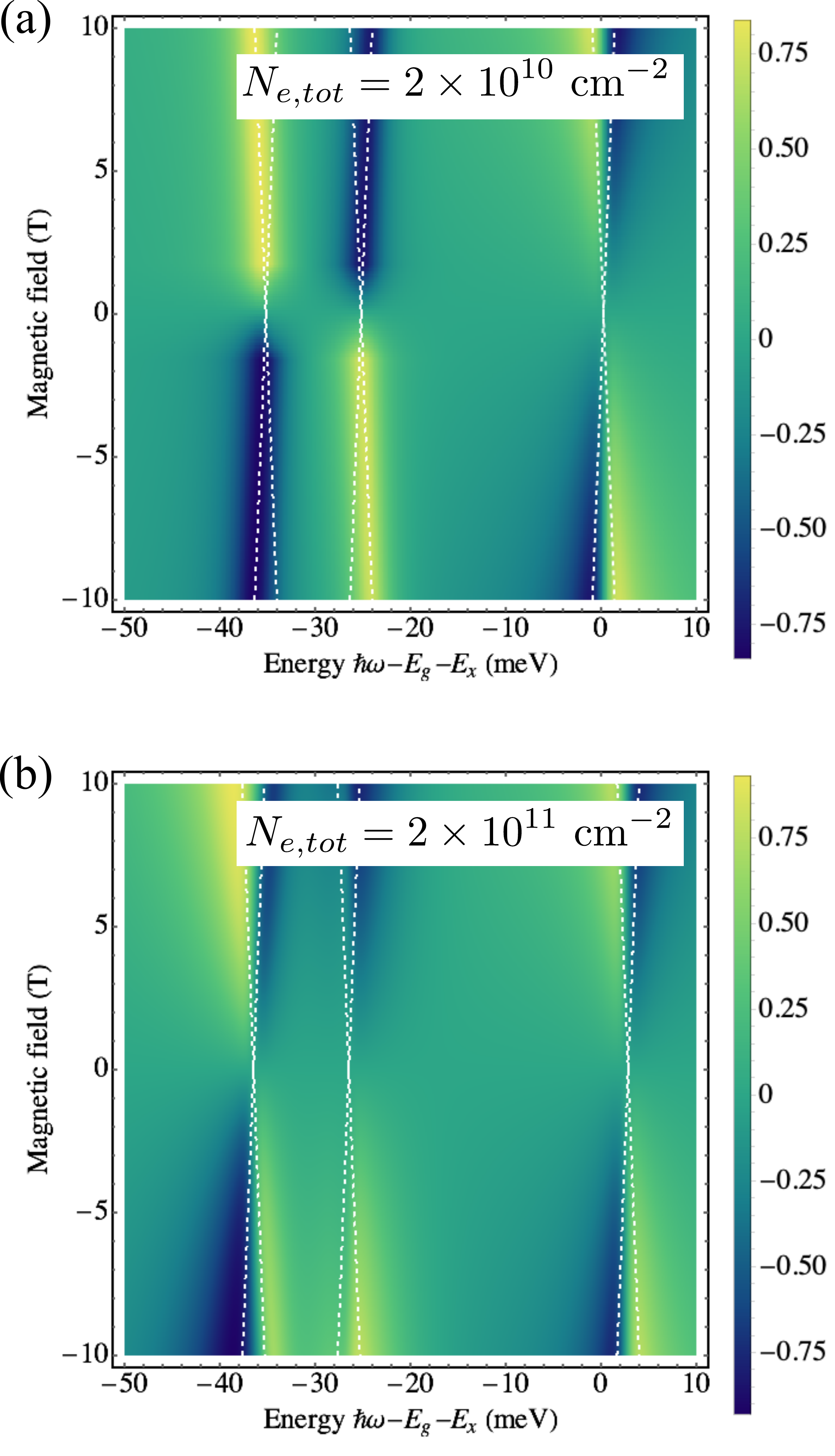}
\caption{\label{fig:Pcirc} False color plot of the circular dichroism of absorption, $P_c(B,\hbar\omega)$ calculated after Eq.~\eqref{Pc} for relatively low doping, $N_{e,tot}=N_{e,\bm K_+} + N_{e,\bm K_-}=2\times 10^{10}$~cm$^{-2}$ (a) and for moderate doping, $N_{e,tot}=2\times 10^{11}$~cm$^{-2}$. Dotted lines show the positions of the Zeeman-split states calculated after Eqs.~\eqref{exc:g:br} and \eqref{opt:trion}. The zero-field positions of the resonances are adjusted with account for the $\propto N_{e,\bm K_\pm} \mathcal D^{-1}$ shifts of the states, Eq.~\eqref{coupled:Gr:1}. The Land\'e factors used in the calculation are as follows $g_v=4$, $g_c=0$, $g_c'=2$. The remaining parameters of calculation are the same as in Fig.~\ref{fig:Ndep}.}
\end{figure}

\subsection{Trion approach}\label{sec:Z:tr}

It follows from Sec.~\ref{sec:osc:tr} that in the course of the trion formation an electron is picked up from the Fermi-sea. Similarly, the trion recombination returns an electron back. Thus, the splitting of the trion transition lines is given by the difference of the Zeeman splitting of the three-particle complex, $X^-_1$ or $X^-_2$,
\begin{align}
\label{trion:g}
\Delta_{Z,tr,1} = \Delta_{Z,c}+\Delta_{Z,c}'-\Delta_{Z,v}, \nonumber \\
\Delta_{Z,tr,2} = \Delta_{Z,c}-\Delta_{Z,c}'-\Delta_{Z,v},
\end{align}
and that of the charge carrier which remains in the system after the recombination. The latter is $\Delta_{Z,c}'$ if the electron remains in the $\bm K_+$-valley or $-\Delta_{Z,c}'$ if the electron remains, Fig.~\ref{fig:bands}. Thus, in both cases, the splitting of the trion line in the optical spectrum is the same as for the neutral exciton:\cite{2053-1583-2-3-034002} 
\begin{equation}
\label{opt:trion}
\Delta_{tr,1} = \Delta_{tr,2} = \Delta_{Z,c} - \Delta_{Z,v} = g_x \mu_B B.
\end{equation}
The effects of Coulomb interaction and bands nonparabolicity which could result in the renormalization of the trion $g$-factor as compared to that of the exciton are disregarded here. Additional renormalization of the $g$-factor related to the fact that the electron is taken and returned from the Fermi sea and having the same origin as the trion energy shift, Eq.~\eqref{delta:k},  is discussed below in Sec.~\ref{sec:Z:FP}.

Importantly, the Zeeman splitting of the resident electrons and corresponding valley polarization, Eq.~\eqref{Pv}, results in the difference of the oscillator strengths of the transitions. Particularly, in accordance with Eq.~\eqref{ftt:2} [cf. Eq.~\eqref{ft}] for transitions active in the $\sigma^+$ polarization, the oscillator strengths of $X_1^-$ and $X_2^-$ are proportional to $N_{e,\bm K_+}$ and $N_{e,\bm K_-}$, respectively. Conversely, for transitions active in the $\sigma^-$ polarization, the oscillator strengths $X_1^-$ and $X_2^-$ are proportional to, respectively, $N_{e,\bm K_-}$ and $N_{e,\bm K_+}$. Thus, at a given circular polarization the oscillator strengths of the intra- and intervalley trions will demonstrate opposite dependence on the magnetic field: One of the trions gains the oscillator strength due to the electron valley polarization, while another one looses it. In the opposite polarization the behavior is opposite.

\subsection{Fermi-polaron approach}\label{sec:Z:FP}

The trion picture outlined above is corroborated by the calculation in the Fermi-polaron approach. Extending Eqs.~\eqref{Pi} and \eqref{coupled:Gr} to allow for the valley degrees of freedom, polarization and Zeeman effect we arrive at the following expressions for the susceptibilites in $\sigma^\pm$ circular polarizations
\begin{equation}
\label{Pi:Z}
\Pi^\pm(\omega) = f_x \mathcal G_x^\pm(\hbar\omega - E_g +E_x;0),
\end{equation}
where the exciton Greens functions read
\begin{widetext}
 %the exciton Greens functions accounting for the multivalley band structure and the Zeeman effect read
\begin{subequations}
\label{coupled:Gr:Z}
\begin{align}
\mathcal G_x^+(\varepsilon) =\frac{1}{\varepsilon - \frac{1}{2}\Delta_{Z,x} + \mathrm i \Gamma - \dfrac{N_{e,\bm K_+} \mathcal D^{-1} E_{tr,1}}{\varepsilon -\frac{1}{2}(\Delta_{Z,tr,1}-\Delta_c') + E_{tr,1}+ \mathrm i \gamma} - \dfrac{N_{e,\bm K_-} \mathcal D^{-1} E_{tr,2}}{\varepsilon -\frac{1}{2}(\Delta_{Z,tr,2}+\Delta_c') + E_{tr,2}+ \mathrm i \gamma}},\\
\mathcal G_x^-(\varepsilon) =\frac{1}{\varepsilon + \frac{1}{2}\Delta_{Z,x} + \mathrm i \Gamma - \dfrac{N_{e,\bm K_-} \mathcal D^{-1} E_{tr,1}}{\varepsilon +\frac{1}{2}(\Delta_{Z,tr,1}-\Delta_c') + E_{tr,1}+ \mathrm i \gamma} - \dfrac{N_{e,\bm K_+} \mathcal D^{-1} E_{tr,2}}{\varepsilon +\frac{1}{2}(\Delta_{Z,tr,2}+\Delta_c') + E_{tr,2}+ \mathrm i \gamma}},
\end{align}
\end{subequations}
\end{widetext}
and 
\[
N_{e,\bm K_\pm} = N_e (1\pm P_v).
\]

Figure~\ref{fig:Pcirc} demonstrates the circular dichroism of absorption
\begin{equation}
\label{Pc}
P_c(B,\hbar\omega) = \frac{\Im\{\Pi^+(\omega)\} - \Im\{\Pi^-(\omega)\}}{\Im\{\Pi^+(\omega)\} + \Im\{\Pi^-(\omega)\}}
\end{equation}
calculated within the Fermi-polaron model. In this calculations we took the set of $g$-factors: $g_v=4$, $g_c=0$, $g_c'=2$, which gives the exciton Land\'e factor $g_x=-4$. We stress that the values of $g$-factors we use are selected here for illustrative purposes, see detailed discussions and microscopic approaches to calculate the Zeeman effect in Refs.~\onlinecite{PhysRevB.95.155406,2020arXiv200202542W,2020arXiv200211646F,2020arXiv200300235D,2020arXiv200211993X}.

In Fig.~\ref{fig:Pcirc} the features in the circular dichroism related to the exciton (repulsive polaron) and trion (attractive polaron) states are clearly seen.
Let us analyze the cases of low and moderate electron densities in more detail. At relatively low electron densities, Fig.~\ref{fig:Pcirc}(a), the trion (attractive polaron) features $X_1^-$ and $X_2^-$ provide significant circular dichroism with opposite signs at the resonances. The Zeeman splitting of the trions is not very prominent here. This is because for the considered set of parameters the complete valley polarization of the resident electrons is achieved at relatively low magnetic field of about $1.65$~T, where the Zeeman splitting of the resonances ($\approx 0.38$~meV) is smaller that the linewidth ($1$~meV). Thus, only one Zeeman component of each trion (attractive polaron) is optically active, namely, the one which requires the electrons remaining in the occupied conduction subband. In contrast, at moderate electron densities, Fig.~\ref{fig:Pcirc}(b), the electron valley polarization is far from complete even at highest magnetic fields. In this case both Zeeman states of each $X_{1,2}^-$ trions are optically active and are visible in the spectra, providing sign-alternating behavior of the circular polarization at each resonance. The exciton oscillator strength just weakly depends on the electron subband occupations [cf. Eq.~\eqref{coupled:Gr:1}] and both Zeeman components of the exciton are present in the circular dichroism spectrum both at the low and moderate electron densities, Fig.~\ref{fig:Pcirc}(a,b). The interplay of the resident electron valley polarization and Zeeman splitting of the excitonic species provides complex dependence of $P_{c}$ on the energy and field shown in Fig.~\ref{fig:Pcirc}. The situation could be even more involved in the case of photoluminescence experiments.\cite{PhysRevLett.121.057402}

It is noteworthy that the valley polarization of the electron gas results in the renormalization of the trion $g$-factor. Indeed, the density-dependent shifts of the attractive polaron energy $\propto N_{e,\bm K_\pm} \mathcal D^{-1}$ [see Eq.~\eqref{coupled:Gr:1} and discussion in Sec.~\ref{sec:osc:FP}] in the presence of magnetic field differ for different Zeeman components and the resulting corrections to the $X^-_{1,2}$ states Zeeman splittings are given by $\pm P_v N_e \mathcal D^{-1}.$ These corrections could be sizeable for moderate electron densities (at small densities these corrections quickly saturate) and could explain observed\cite{Srivastava:2015a,PhysRevLett.114.037401,2053-1583-2-3-034002} differences between the bright exciton and trion $g$-factors. While these corrections are straightforwardly derived in the Fermi-polaron approach, they can be also estimated in the trion approach if one takes into account the fact that for the trion formation the electron is picked up from the Fermi sea, which results in the shift of the trion resonance [cf. Eq.~\eqref{delta:k}]. We stress that at low electron densities where both approaches merge this contribution to $g$-factor could be important only at small magnetic fields. Again, the key features of the trion (attractive polaron) fine structure can be evaluated both in the trion and Fermi-polaron models with the same result provided that the resident electron density is low enough and the conditions~\eqref{conditions} are satisfied.

\section{Photoluminescence at non-resonant excitation}\label{sec:PL}

Above we discussed resonant optical properties of TMDC MLs in the spectral range of neural and charged excitons. Particularly, $-\Im{\Pi}$ given by Eq.~\eqref{Pi}, provides the absorption spectrum via the exciton Greens function. \add{An alternative} experimental approach to study the Coulomb-bound electron-hole complexes is to observe photoluminescence (or resonant light scattering) under non-resonant excitation where the electron-hole pairs or excitons are formed with high excess energy and eventually relax to the low-energy radiative states. Below we briefly discuss the trion formation process and photoluminescence effect from the trion and Fermi-polaron viewpoints. In this section we disregard the complex band structure of the TMDC MLs.

\subsection{Trion approach}\label{sec:PL:tr}

Here we analyze the formation of the trions from excitons in 2D TMDC where the energy \add{difference} between the exciton and the trion states is close to the energy of the optical phonon, $\hbar\Omega$. We assume that the main process governing the trion photoluminescence is related to the trion formation and its subsequent radiative recombination, leaving out the discussion of the thermalization issues.\cite{zipfel:trions:marina} We develop the model of the capture of the electron by exciton to form a trion following the general approach in Ref.~\onlinecite{abakumov_perel_yassievich}. The exciton-electron interaction is modelled as a zero-radius potential, Sec.~\ref{sec:model}. Free electron wavefunction with the in-plane wavevector $\bm k$ reads
\begin{multline}
\label{free}
\Phi_{\bm k}(\bm \rho) = e^{\mathrm i \bm k \bm \rho} + f_{k} \sqrt{\frac{\pi k}{2}} \mathrm i {\rm H}_0^{(1)} (k\rho) \approx
e^{\mathrm i \bm k \bm \rho} + f_{k}  \frac{e^{\mathrm i k \rho}}{\sqrt{-\mathrm i \rho}}.
\end{multline}
Here, for convenience, we used the scattering amplitude $f_k$ in the coordinate normalization:\cite{ll3_eng}
\begin{equation}
\label{fk:te}
f_k = - \sqrt{\frac{2\pi}{k}} \mathcal D T(\varepsilon), \quad \varepsilon=\frac{\hbar^2 k^2}{2m},
\end{equation}
and $T(\varepsilon)$ is given by Eq.~\eqref{scattering:ampl}. Note that in our model the interaction takes place only in the channel with the angular momentum component $l_z=0$.
It is instructive to check the orthogonality relation between the bound [electron-in-trion, Eq.~\eqref{wave:tr:bound}] and free-electron [Eq.~\eqref{free}] states, which is necessary to properly calculate the capture rate:
\begin{multline}
\label{orth}
\int d\bm \rho \Phi_0(\rho) \Phi_{\bm k}(\bm \rho)\\ = \left\{\frac{2\sqrt{\pi}\ae}{k^2+\ae^2} + 2f_k\frac{\sqrt{2k}\ae}{k^2+\ae^2}\ln{\left(\mathrm i \frac{k}{\ae} \right)}
\right\}=0.
\end{multline}
Here $\ae=a_{tr}^{-1}$. Making use of the explicit form of $f_k=-\sqrt{\pi/2k}\ln^{-1}(\mathrm i \ae/k)$, Eqs.~\eqref{scattering:ampl:1} and \eqref{fk:te}, one can see that the expression in curly brackets of Eq.~\eqref{orth} is identically zero. 

Under non-resonant excitation the trion is formed when exciton captures the resident electrons and emits optical phonon to ensure the energy conservation. This process can be considered as a trapping of the electron by the effective potential well created by the exciton accompanied by the phonon emission.

The matrix element of the optical phonon emission which couples free and bound states can be written in the simplest approximation as 
\begin{equation}
\label{Mkq}
M_{\bm k}^{\bm q} = C_0(q) \int d\bm \rho \Phi_0(\rho) e^{-\mathrm i \bm q\bm \rho} \Phi_{\bm k} (\bm \rho), 
\end{equation}
with  $\bm q$ being the phonon wavevector, $C_0(q)$ being a parameter [see Refs.~\onlinecite{7496798,PhysRevB.94.085415} for the explicit form of the F\"ohlich interaction in 2D systems]. Note that at $\bm q=0$ the matrix element~\eqref{Mkq} vanishes due to the orthogonality of the wavefunctions.
Now we are able to calculate the trion formation rate (per exciton with the given energy $E_k$) making use of the Fermi's golden rule:\cite{abakumov_perel_yassievich}
\begin{equation}
\label{nu}
\nu_{x}(E_k) = \frac{2\pi}{\hbar}  N_e \sum_{\bm q} |M_{\bm k}^{\bm q} |^2 \delta(E_k -\hbar\Omega + E_{tr}),
\end{equation}
where the $\delta$-function describes the energy conservation, and we neglected the trion dispersion. Assuming that $C_0(q)$ weakly depends on $q$ and replacing it by its $q=0$ value we can perform the summation in Eq.~\eqref{nu} over the phonon wavevector with the result
\begin{equation}
\label{nu:1}
\nu_{tr}(E_k) = \frac{2\pi}{\hbar} |C_0|^2 \mathfrak I_k N_e \delta(E_k -\hbar\Omega + E_{tr}),
\end{equation}
where 
\begin{equation}
\label{Ik:def}
\mathfrak I_k = \int d\bm \rho |\Phi_0(\rho) \Phi_{\bm k}(\bm \rho)|^2.
\end{equation}

The trion generation rate is given by
\begin{equation}
\label{nu:trion:gen}
W_{tr} = \sum_{\bm k} \nu_x(E_k) n_x(E_k),
\end{equation}
where $n_x(E)$ is the exciton distribution function formed as a result of the non-resonant excitation. The decay rate of the trion is $2\gamma/\hbar$. Thus, the steady-state trion population is $ \hbar W_{tr}/(2\gamma)$. Correspondingly, the trion photoluminescence spectrum can be presented as [cf. Eq.~\eqref{Pi:trion}]:
\begin{equation}
\label{trion:PL}
I(\hbar\omega) \propto \frac{1}{\pi}\frac{\gamma}{(\hbar\omega - E_g + E_x + E_{tr})^2+\gamma^2} \frac{\hbar W_{tr}}{2\gamma}.
\end{equation}
This treatment agrees with results of the approach developed in Ref.~\onlinecite{PhysRevLett.122.217401} where the processes of exciton recombination via capture to the localized electron centers have been considered. Equation~\eqref{trion:PL} is valid provided that phonon-induced trion dissociation rate, $W_{diss} \propto W_{tr} \exp{(-\hbar\Omega/k_B T)}$, is slow as compared with its decay rate $2\gamma/\hbar$. In Eq.~\eqref{trion:PL} we neglected the energy shifts and recoil effects [cf. Eqs.~\eqref{Pi:trion} and \eqref{delta:k}], see Refs.~\onlinecite{PhysRevB.62.8232,Manassen:96} for detail. As before, the latter approximation is strictly justified at $E_F \lesssim \gamma$.

\subsection{Fermi-polaron approach}\label{sec:PL:FP}

In the Fermi-polaron approach the trion generation and photoluminescence can be readily calculated using the Keldysh diagram technique following Refs.~\onlinecite{deych:075350,Averkiev:2009aa}. We introduce the Greens function
\begin{equation}
\label{GK}
\mathcal G_x^{-+}(\varepsilon,\bm k) = n(\varepsilon) \left[\mathcal G_x^{*}(\varepsilon,\bm k) - \mathcal G_x(\varepsilon,\bm k)\right],
\end{equation}
which accounts for the non-equilibrium distribution of the quasi-particles $n(\varepsilon)$. The remaining Greens functions in the Keldysh technique in the lowest order in $n(\varepsilon)$ read: $\mathcal G_x^{--} = \mathcal G_x$, $\mathcal G_x^{++} = - \mathcal G_x^{*}$. The photoluminescence spectrum is proportional to 
\begin{equation}
\label{FP:PL}
I(\hbar\omega) \propto f_x \Im \{\mathcal G_x^{-+}(\hbar\omega - E_g - E_x,0)\}
\end{equation}
If the excitonic subsystem were in thermal quasi-equilibrium, $n(\varepsilon) \propto \exp{[(\mu_c-\varepsilon)/k_B T]}$. Below, like in Sec.~\ref{sec:PL:tr} we focus on the non-equilibrium situation where the photolumenescence of Fermi-polarons (trions) is controlled by the optical phonon-induced transitions.

Following the rules of the Keldysh technique we evaluate the $\mathcal G_x^{-+}(\varepsilon,0)$ accounting from the phonon-assisted transitions from the higher-energy excitonic states in the first order:
\begin{multline}
\label{dG-+}
\delta \mathcal G_x^{-+}(\varepsilon,0) = - \mathcal G_{x}^{--}(\varepsilon,0) \mathbb \Sigma^{-+} \mathcal G_{x}^{++}(\varepsilon,0) \\
=\frac{1}{\hbar}\sum_{\bm k,\bm q} |M^{\bm q, eff}_{\bm k}|^2 \Im\{\mathcal G_x^{-+}(\varepsilon+\hbar\Omega,\bm k)\}  \left|\mathcal G_x(\varepsilon,0)\right|^2
\end{multline}
Here expressed the self-energy $\mathbb \Sigma^{-+}$ via the Greens function $\mathcal G_x^{-+}$ and the effective matrix element (vortex) of exciton-phonon interaction $M^{\bm q, eff}_{\bm k}$, which should  be calculated with allowance for the exciton-electron interaction, see Appendix~\ref{sec:app}. This approximation corresponds to the neglect of the phonon-induced transitions to the higher energies.
At $\varepsilon \approx -E_{tr}$ the Greens function $\mathcal G_x^{-+}(\varepsilon+\hbar\Omega,\bm k)$ can be replaced by the $2\pi\mathrm i \delta(E_k-\hbar\Omega +E_{tr}) n(E_k)$, while in evaluation of $|\mathcal G_x(\varepsilon,0)|^2$ one has to keep the contribution linear in $N_e$ resulting from the interference of the first and second terms in Eq.~\eqref{coupled:Gr:1}. Neglecting the term $N_e \mathcal D_{e}^{-1}$ in the denominator, we arrive at Eq.~\eqref{trion:PL}, with 
%the following expression for the trion generation rate
\begin{equation}
\label{nu:trion:gen:1}
W_{tr} = \frac{N_e}{\mathcal D E_{tr}} \frac{2\pi}{\hbar} \sum_{\bm k,\bm q} |M^{\bm q, eff}_{\bm k}|^2  n(E_k) \delta(E_k - \hbar\Omega + E_{tr}).
\end{equation}

To establish the  agreement of the approaches we need to calculate $M^{\bm q, eff}_{\bm k}$ and compare Eq.~\eqref{nu:trion:gen:1} with the result of the Fermi's golden rule, Eq.~\eqref{nu:trion:gen}. The calculations presented in Appendix~\ref{sec:app} show that
\begin{equation}
\label{M:eff:a}
|M^{\bm q, eff}_{\bm k}|^2 = \mathcal D E_{tr} |M_{\bm k}^{\bm q}|^2.
\end{equation}
Therefore, Eqs.~\eqref{nu:trion:gen} and \eqref{nu:trion:gen:1} are consistent, and both approaches provide the same result. Thus, it is a matter of convenience to select the approach to calculate the photoluminescence, provided that Eqs.~\eqref{conditions} are fulfilled and more complex processes of trion-electron scattering can be neglected.

\section{Conclusion}

To conclude, we have demonstrated by several examples that optical properties of charged excitons in transition metal dichalcogenide monolayers can be described both in the trion and in the Fermi-polaron approach, provided the following hierarchy of energy is fulfilled: The exciton binding energy exceeds by far the trion binding energy which, in its turn, exceeds the electron Fermi energy. Direct analysis of (i) the optical transition oscillator strengths, (ii) the spectrum fine structure and Zeeman effect, as well as  (iii)  the photoluminescence demonstrates that these effects can be adequately described both in the trion and in the Fermi-polaron pictures taking into account simplifications behind each approach.

There are several interesting and important problems to be addressed in the future. On the one hand, it is desirable to explore the high electron density regime where the Fermi energy of the charge carriers is comparable of exceeds the trion binding energy. Also, the description of the trion/Fermi-polaron transport properties, e.g., the effect of the photoconductivity [cf. Ref.~\onlinecite{PhysRevX.9.041019}] in the spectral range of charged excitons resonance is an interesting and important problem to be addressed in future. It is likewise important to search for the experimentally accessible situations where the exciton-electron correlations in two-dimensional semiconductors are so strong that the simplified approaches outlined above become inapplicable.

\begin{acknowledgments}
I am grateful to A. Chernikov, A. Imamoglu, D. Reichman, M.A. Semina, T. Smolenski, and R. Schmidt for valuable discussions.
This work was partially supported by the Russian Science Foundation (project \#~19-12-00051).
\end{acknowledgments}

\section*{DATA AVAILABILITY STATEMENT}

The data that support the findings of this study are available from the corresponding author upon reasonable request.

\appendix

\section{Exciton self-energy}\label{sec:app:self}

Let us discuss in more detail the approximations behind Eq.~\eqref{exciton:self} which states $\Sigma(\varepsilon,\bm k) = T(\varepsilon) N_e$. Such expression corresponds to the Hartree type of the self-energy where the correlations and self-consistent effects are disregarded. If the exciton-electron interaction were weak and described by the matrix element $V_0$, and the perturbation theory in the electron-exciton interaction were applicable, this expression would correspond to the first-order perturbation theory
\begin{equation}
\label{sigma:first:order}
\Sigma^{(1)} = V_0 N_e.
\end{equation}
Following general arguments~\cite{ll9_eng} one could replace in Eq.~\eqref{sigma:first:order} the perturbation matrix element $V_0$ by the full scattering amplitude $\mathcal T(\varepsilon)$ in order to account for the main contributions due to the higher orders in $V_0$ and arrive at Eq.~\eqref{exciton:self}.

Although being the simplest possible approximation, Eq.~\eqref{exciton:self} captures the key effects and allows for the fully analytical solution of the problem. More sophisticated treatment of the problem\cite{suris:correlation,PhysRevA.84.033607,PhysRevLett.65.1032,PhysRevB.45.12419,PhysRevB.98.235203,PhysRevA.85.021602,PhysRevX.9.041019} demonstrate that qualitative differences could appear. The state-of-the-art approach\cite{PhysRevA.85.021602,PhysRevX.9.041019} is to solve self-consistently the set of equations
\begin{subequations}
\label{sigma:self}
\begin{equation}
\Sigma(\varepsilon,\bm k) = S^{-1}\sum_{\bm p} n_{ \add{\bm p}} T\left(\varepsilon + \frac{\hbar^2 p^2}{2m_e} , \bm k + \bm p \right),\label{sigma:self:A}
\end{equation}
\begin{multline}
T^{-1} (\varepsilon,\bm k) = -\mathcal D \ln{\left(\frac{\bar E}{E_{tr}}\right)}  \\\add{-} S^{-1}\sum_{\bm p} (1-n_{\bm p})\mathcal G\left(\varepsilon - \frac{\hbar^2 p^2}{2m_e} ;\bm k -\bm p\right),\label{T:gen}
\end{multline}
\end{subequations}
with $S$ being the normalization area and $n_{\bm p}$ being the electron distribution function. 

Note that neglecting $n_{\bm p}$ in Eq.~\eqref{T:gen}, which could be justified for $E_F \ll E_x, E_{tr}$ where the main part of the integration involves empty states, substituting bare exciton Greens function, and taking $\bm k=0$ we  arrive to Eq.~\eqref{scattering:ampl:1}. Furthermore, if in integration in Eq.~\eqref{sigma:self:A} we disregard the electron dispersion we arrive at Eq.~\eqref{exciton:self} (for full analytical solution for $T$ see Refs.~\onlinecite{PhysRevA.85.021602,PhysRevB.95.035417}). This approximation, however, overestimates exciton-electron interaction and, particularly, the polaron repulsion parameter $\delta$ in Eq.~\eqref{shift}. This is because at $\varepsilon \approx -E_{tr}$ the scattering amplitude has a pole and strongly depends on its arguments.

One possible extension is to introduce an additional parameter of the theory $0<\xi<1$ which phenomenologically takes into account self-consistent effects and reduce the exciton self-energy as
\begin{equation}
\label{exciton:self:xi}
\Sigma(\varepsilon;\bm k) = \xi T(\varepsilon) N_e.
\end{equation}
In fact, this is equivalent to artificial decrease of the electron density in the expressions presented in the main text.

\section{Calculation of the effective matrix element of Fermi-polaron interaction with phonons}\label{sec:app}

In order to determine $M_{\bm k}^{\bm q,eff}$ we calculate the self-energies $\mathbb \Sigma^{--}(\varepsilon, \bm k)$ describing the damping of excitons with $\varepsilon \approx 0$ due to the phonon emission, and $\mathbb \Sigma^{-+}(\varepsilon,\bm k)$, describing the generation of trions with $\varepsilon \approx - E_{tr}$ due to the phonon absorption. We need full exciton Green's function $\mathcal G_{x}(\varepsilon; \bm k, \bm k')$ which depends on the ``initial'' and ``final'' wavevectors of the excitons. For the case of an exciton interacting with a single electron it reads\cite{suris:correlation}
\begin{equation}
\label{G:full}
G_x^{(1)}(\varepsilon; \bm k, \bm k') =\delta_{\bm k,\bm k'}G_\varepsilon(\bm k) + \frac{T(\varepsilon)}{S}G_\varepsilon(\bm k)G_\varepsilon(\bm k') ,  
\end{equation}
with $T(\varepsilon)$ being the scattering amplitude, Sec.~\ref{sec:model}, $S$ the normalization area, and 
\[
G_{\varepsilon}(\bm k)=  \frac{1}{\varepsilon - E_{\bm k} + \mathrm i \Gamma}.
\]
The self-energy of exciton related to the  phonon emission can be recast as 
\begin{multline}
\mathbb \Sigma^{--} = \sum_{\bm q} |C_0(q)|^2 \Im\bigl\{\sum_{\bm k',\bm p',\bm p_1',\bm p_1} \delta_{\bm p_1, \bm p_1' + \bm q}  \delta_{\bm p', \bm k - \bm q} \\
\times \left[ \delta_{\bm k,\bm k'} +\frac{T(\varepsilon)}{S}G_\varepsilon(\bm k')\right]\left[ \delta_{\bm p,\bm p_1} +\frac{T(\varepsilon)}{S}G_\varepsilon(\bm p_1)\right]\\
\times  G_x^{(1)}(\varepsilon- \hbar\Omega, \bm p', \bm p_1')\bigr\}.
\end{multline}
At $\varepsilon - \hbar\omega \approx - E_{tr}$ it is sufficient to account for the term $\propto T(\varepsilon- \hbar\omega)$ in $G_x^{(1)}(\varepsilon- \hbar\Omega, \bm p', \bm p_1')$. Furthermore, allowing for the trion damping we present $T(\varepsilon- \hbar\Omega) = -\mathrm i \pi \mathcal D^{-1} E_{tr} \delta(\varepsilon- \hbar\Omega + E_{tr})$ and perform summation over $\bm k', \bm p', \bm p_1', \bm p_1$ transforming the Greens functions to the real space. Taking into account the finite density of electrons (replacement $S^{-1} \to N_e$) and using Eqs.~\eqref{free}, \eqref{orth}, and \eqref{Mkq} we arrive at
\begin{subequations}
\label{Sigma--}
\begin{equation}
\mathbb \Sigma^{--}(\varepsilon,\bm k) = -\pi N_e \sum_{\bm q} |M_{\bm k}^{\bm q}|^2\delta(\varepsilon- \hbar\Omega + E_{tr}).
\end{equation}
On the other hand, 
\begin{equation}
\mathbb \Sigma^{--} = \sum_{\bm k,\bm q}  |M_{\bm k}^{\bm q, eff}|^2 \Im\{ \mathcal G_x^{--}(\varepsilon- \hbar\Omega,\bm k)\}.
\end{equation}
\end{subequations}
Equations~\eqref{Sigma--} are consistent provided that 
\begin{equation}
\label{M:eff:a}
|M^{\bm q, eff}_{\bm k}|^2 = \mathcal D E_{tr} |M_{\bm k}^{\bm q}|^2.
\end{equation}
Similar transformations allow us to present ($\varepsilon \approx - E_{tr}$)
\begin{subequations}
\label{Sigma-+}
\begin{equation}
\mathbb\Sigma^{-+}(\varepsilon,0) = 2\pi \sum_{\bm k,\bm q} n(E_k) |M^{\bm q}_{\bm k}|^2 \delta(E_k - \hbar\Omega - \varepsilon),
\end{equation}
which with allowance for Eq.~\eqref{M:eff:a} is consistent with the definition [Eq.~\eqref{dG-+}] 
\begin{equation}
\mathbb\Sigma^{-+}(\varepsilon,0) = \sum_{\bm k,\bm q} |M^{\bm q, eff}_{\bm k}|^2 \Im\{\mathcal G_x^{-+}(\varepsilon+\hbar\Omega,\bm k)\}.
\end{equation}
\end{subequations}


%\bibliography{/Users/misha/Work/Coherent/Bibliography/all-1}


\end{document}